\documentclass[useAMS,usenatbib]{mnras}

\usepackage{hyperref}

\usepackage{epsfig}
\usepackage{epstopdf}
\usepackage{graphicx,amssymb,longtable,lscape} 
\usepackage[fleqn]{amsmath}
\usepackage{textcomp}
\usepackage{color}
\usepackage{multirow}
\usepackage{float}

\usepackage[english]{babel}
\usepackage{array}
\newcolumntype{L}[1]{>{\raggedright\let\newline\\\arraybackslash\hspace{0pt}}m{#1}}
\newcolumntype{C}[1]{>{\centering\let\newline\\\arraybackslash\hspace{0pt}}m{#1}}
\newcolumntype{R}[1]{>{\raggedleft\let\newline\\\arraybackslash\hspace{0pt}}m{#1}}

\def\apj{\mbox{ApJ}}
\def\apjl{\mbox{ApJL}}
\def\apjs{\mbox{ApJS}}
\def\aaps{\mbox{A\&AS}}
\def\mnras{\mbox{MNRAS}}
\def\aj{\mbox{AJ}}
\def\araa{\mbox{ARA\&A}}
\def\pasp{\mbox{PASP}}
\def\nat{\mbox{Nature}}
\def\aap{\mbox{A\&A}}

\title[Extreme star formation rates in quasar hosts]
{Extreme star formation events in quasar hosts over ${\bf0.5<\textit{z}<4}$}

\author[Pitchford et al.]{L.~K. Pitchford$^{1,2}$, E. Hatziminaoglou$^1$, A. Feltre$^3$, D. Farrah$^2$, C. Clarke$^4$,  
\newauthor
K.A. Harris$^{2,5}$, P. Hurley$^{4}$, S. Oliver$^4$, M. Page$^6$, L. Wang$^7$\\
$^{1}$ESO, Karl-Schwarzschild-Str. 2, 85748 Garching bei M\"unchen, Germany\\
$^{2}$Department of Physics, Virginia Tech, Blacksburg, VA 24061, USA\\
$^{3}$Sorbonne universit{\'e}s, UPMC-CNRS, UMR7095, Institut d'Astrophysique de Paris,
F-75014, Paris, France\\
$^{4}$Astronomy Centre, Dept. of Physics \& Astronomy, University of Sussex, Brighton BN1 9QH, UK\\
$^{5}$Instituto de Astrof{\'i}sica de Canarias, C/via L{\'a}ctea s/n, E-38205 La Laguna, Tenerife, Spain \\
$^{6}$Mullard Space Science Laboratory, University College London, Holmbury St Mary, Dorking, 
Surrey RH5 6NT, UK\\
$^{7}$SRON Netherlands Institute for Space Research, Landleven 12, NL-9747 AD, Groningen, The Netherlands\\
}

\begin{document}

\pagerange{\pageref{firstpage}--\pageref{lastpage}} \pubyear{2016}

\maketitle

\label{firstpage}

\begin{abstract}
We explore the relationship between active galactic nuclei and star formation in a sample of 513 optically luminous type 1 quasars up to redshifts of $\sim$4 hosting extremely high star formation rates (SFRs). The quasars are selected to be individually detected by the \textit{Herschel} SPIRE instrument at $>$\,3$\sigma$ at 250\,$\mu$m, leading to typical SFRs of order of 1000\,M$_{\odot}$yr$^{-1}$. We find the average SFRs to increase by almost a factor 10 from $z\sim0.5$ to $z\sim3$, mirroring the rise in the comoving SFR density over the same epoch. However, we find that the SFRs remain approximately constant with increasing accretion luminosity for accretion luminosities above 10$^{12}$\,L$_{\odot}$. We also find that the SFRs do not correlate with black hole mass. Both of these results are most plausibly explained by the existence of a self-regulation process by the starburst at high SFRs, which controls SFRs on time-scales comparable to or shorter than the AGN or starburst duty cycles. We additionally find that SFRs do not depend on Eddington ratio at any redshift, consistent with no relation between SFR and black hole growth rate per unit black hole mass. Finally, we find that high-ionisation broad absorption line (HiBAL) quasars have indistinguishable far-infrared properties to those of classical quasars, consistent with HiBAL quasars being normal quasars observed along a particular line of sight, with the outflows in HiBAL quasars not having any measurable effect on the star formation in their hosts. 
\end{abstract}


\section{Introduction}\label{intro}
Nuclear activity and star formation in galaxies are observed to often coexist across all redshifts \citep{farrah03, alexander05, shi09, hernan09, hatzimi10, mainieri11, lamassa13,harris16,alag16}, up to extremely high active galactic nuclei (AGN) and starburst luminosities \citep[e.g.][]{gen98,carilli01,omont01,farrah02,efst14,magdis14,rosen15}. Moreover, there is a tight correlation between the stellar velocity dispersion in the bulge and the mass of the supermassive black hole (BH) residing in the centre for nearby quiescent galaxies (e.g. \citealt{magorrian98,ferrarese00,tremaine02,haering04}). These observations suggest that there exists a deep connection between stellar and black hole mass assembly events in galaxies. 

The nature of, and connection between, star formation and AGN activity in galaxies is unclear. For starbursts, the majority of star-forming systems lie on a `main sequence' whose mean star formation rate (SFR) rises with redshift, from about 10\,M$_\odot \mathrm{yr}^{-1}$ at $z=0.5$ to roughly 100\,M$_\odot \mathrm{yr}^{-1}$ at $z=2$ \citep[as in e.g.][]{elb11,rodig11,schreiber15}. The origin of this main sequence is however still debated, as is the trigger for star formation as a function of a galaxy's position relative to the main sequence. For AGN, it is difficult to complete a robust census of AGN from surveys, since they are obscured by dust and gas for a significant fraction of their lives, with this fraction possibly dependent upon both luminosity and redshift \citep[e.g.][]{mart05}. It is also unclear if star formation events and AGN activity can directly affect one another. A direct relation is motivated by models for galaxy assembly to improve consistency between predictions and observations, most often via `quenching' of star formation by an AGN \citep[e.g.][]{bower06, croton06, booth09, fabian12}. However, observational studies of quenching remain inconclusive. Indirect manifestations of feedback, such as large molecular gas outflows \citep[e.g.][]{feruglio10,spo13} and powerful AGN-driven winds \citep[e.g.][]{perna15}, are found routinely (see also e.g. \citealt{farrah09,bridge13,spo13}), but studies that claim a causal relation are rarer \citep{farrah12,page12}, and sometimes controversial \citep[e.g.][]{harrison14}.

As a result, the scaling relations between star formation and AGN activity across their respective duty cycles remain uncertain. Some authors find a scaling between SFR and AGN luminosity \citep[e.g][]{ima11,Young2014,delv15}, while others do not \citep[e.g.][]{shao10,mullaney12,harrison14,ma15,stanley15}. Recently, \citet{harris16} have shown that a correlation between SFR and AGN luminosity may only exist over certain AGN luminosity, SFR and redshift ranges. 

An insightful way to study the connections between star formation and AGN activity in the context of galaxy assembly events is to examine their scaling relations in populations that signpost specific regions of the AGN luminosity and SFR parameter space. One such population are optically luminous type 1 quasars that host luminous, off-main sequence starbursts. This population is straightforward to find in optical surveys, corresponds to a specific phase in the AGN duty cycle, and may signpost the extremes of the starburst duty cycle. As such, they may illustrate how the processes that convert free baryons to stellar and BH mass may change at the most extreme luminosities. They are also an excellent population in which to search for the initial stages of AGN feedback. 

A related population within which it would appear intuitive to search for evidence of AGN feedback is that of the Broad Absorption Line (BAL) quasars \citep{lynds67,turnshek84}. The BAL quasars possess broad P-Cygni-like absorption features in their ultraviolet spectra that are blue-shifted with respect to the nominal wavelength of the emission line. These features may stem from outflows from the quasars and may further be associated with high mass-loss rates \citep{dekool02, chartas03}. These outflows could arise in two ways; the outflows could be a random process, present for only a random fraction of the quasar lifetime and/or over certain viewing angles \citep[e.g.][]{elvis00}. The observation of BALs in only 10-15 per cent of quasars \citep[e.g.][]{hewett03} then arises from a suitable combination of these two factors. Second, BALs may signpost outflows that occur only at a particular point in a quasar's lifetime, most likely `young' objects recently (re)fuelled by mergers that also trigger star formation in their hosts. It is this second scenario that would mean that BAL quasars may signpost quasar-mode feedback. This may manifest itself via BAL quasars having different far-infrared (FIR) properties, on average, to those of ordinary quasars. 

In this paper, we undertake such a study, exploring the relationship between star formation and AGN activity in optically luminous quasars over $0.5<z<4$ that host extremely high SFRs, from 40$-$4000\,M$_\odot$yr$^{-1}$, placing these hosts approximately one dex above the star formation main sequence. To do so, we start with quasars in the Sloan Digital Sky Survey (SDSS) that also lie within extragalactic survey fields covered by the Spectral and Photometric Imaging Receiver \citep[SPIRE;][]{griffin10} instrument onboard \textit{Herschel} \citep{pilbratt10}. We then restrict ourselves to those quasars that are individually detected by \textit{Herschel}. We infer SFRs from the \textit{Herschel} data, and compare these to the properties of the AGN, as measured from the SDSS catalogue data. We also examine the properties of BAL quasars to see if their properties show evidence for AGN feedback. 

The paper is structured as follows. The SDSS and {\it Herschel} data, as well as the matched quasar catalogue, are described in Sec. \ref{sec:data}. Further, Sec. \ref{sec:analysis} describes the spectral energy distribution (SED) fitting, the SFR derivation and uncertainties and discusses the way in which the current study complements previous works in terms of samples and methodology. Sec. \ref{sec:results_full} presents the SFRs in the full quasar sample and in the BAL quasar sub-sample separately, as a function of the quasars' intrinsic properties, namely accretion luminosity, BH mass and Eddington ratio. Lastly, Sec. \ref{sec:discussion} discusses our results and places them in a greater context. Throughout this work, we assume $H_0 = 72 \hspace{0.1cm} \textnormal{km} \hspace{0.1cm} \textnormal{s}^{-1} \hspace{0.1cm} \textnormal{Mpc}^{-1}$, $\Omega_\Lambda = 0.7$ and $\Omega_\textnormal{M}=0.3$.

\begin{table*}
	\begin{center}
		\begin{tabular}{l C{2cm} crrrrrrr}
			\hline
			&&& \multicolumn{3}{c}{SDSS quasars}  && \multicolumn{3}{c}{{\it Herschel}/SDSS quasars} \\
			\cline{4-6}
			\cline{8-10}
			\noalign{\vskip0.025cm}
			Field & Overlapping Region (deg$^2$) && Total & In DR7 & In DR10 && Total & In DR7 & In DR10\\
			\hline	
			\vspace{-0.25cm}
			&&&&&&&&\\
				HerMES & 60  && 1,293 & 586 & 739 && 83 & 43 & 44\\
				HerS       & 79  && 4,524 & 2,154 & 2,779 && 212 & 150 & 90\\
				HeLMS  & 210 && 8,827 & 2,488 & 6,796 && 218 & 92 & 144\\ 
			\hline
				
		\end{tabular}
		\caption{Quasar detection statistics by survey. The overlapping region is the area of each \textit{Herschel} field that is also covered by SDSS. The SDSS detections are those quasars that lie in their respective \textit{Herschel} fields. The joint detections columns represent those SDSS quasars with SPIRE detections above 3$\sigma$ at 250$\micron$. We further describe these objects by the data release in which they are found: DR7 or DR10. The sum of the DR7 and DR10 columns is always higher than the columns showing the totals, as the latter only consider unique entries. The SDSS detections columns represent those SDSS quasars located in each respective \textit{Herschel} field, while the joint detections are the quasars with 250\,$\micron$ detections at $>$3$\sigma$.}
		\label{tab:jointDetections}
	\end{center}
\end{table*}

\begin{figure}
\begin{center}
\vspace{-.2cm}
\includegraphics[width=8cm]{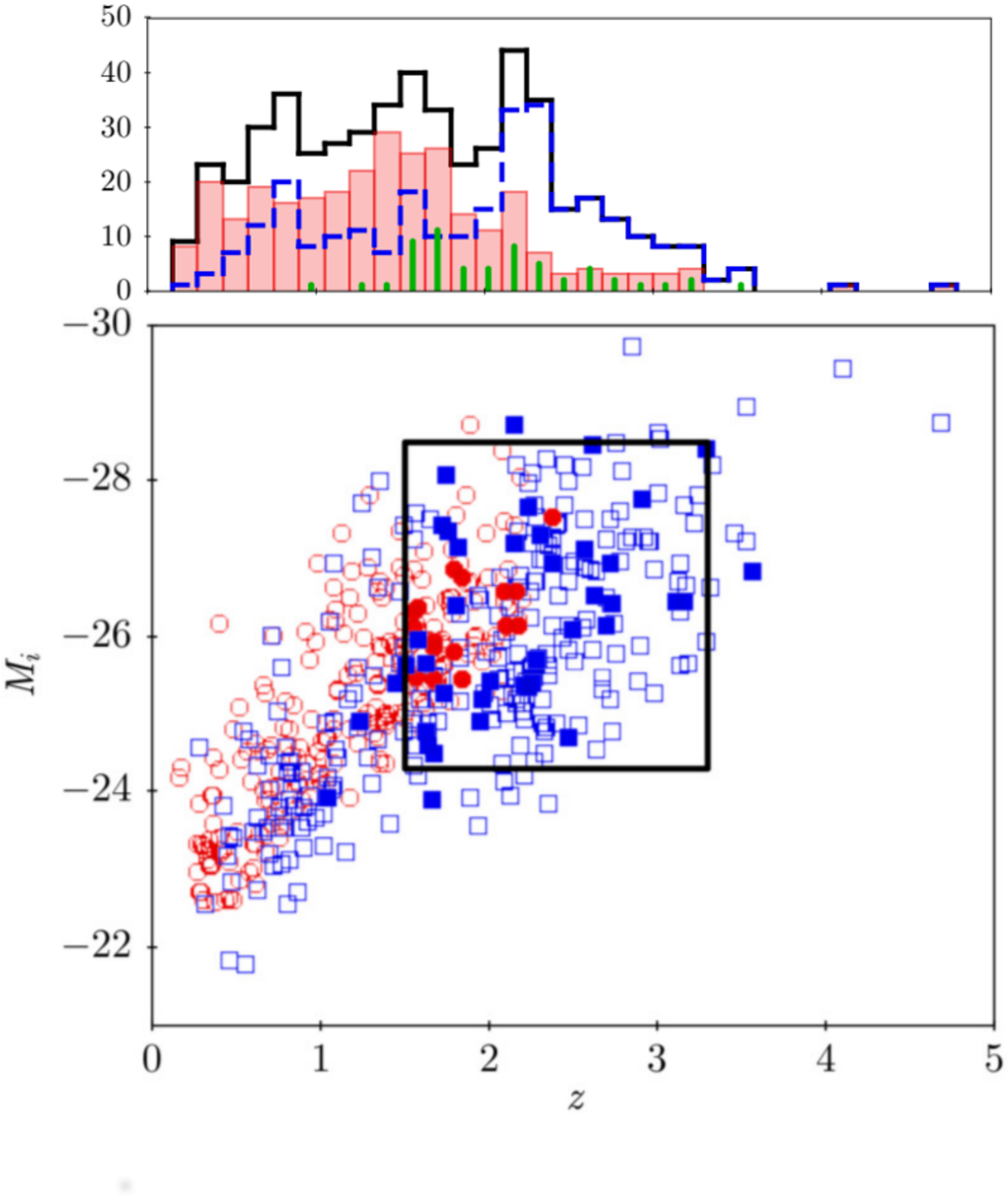}
\vspace{-2.5cm}
\caption{Absolute \textit{i}-band magnitude versus redshift. The red circles represent those objects included in SDSS DR7, while the blue squares are those objects in DR10.
The filled and open shapes differentiate between the BAL and non-BAL samples, respectively (see Section \ref{sec:BAL}).  To characterise these BAL and non-BAL subsets we use the objects that lie within the black box, with its limits defined by 1.5\,$\le$\,\textit{z}\,$\le $\,3.3 and $-28.5$\,$\le$\,$M_i$\,$\le$\,$-24.3$, in which most of the BAL quasars lie. The redshift histograms are shown at the top, with the filled red, dashed blue and solid black showing the DR7, DR10 and full sample of objects, respectively. The green spikes represent the redshift distribution of the BAL quasars.}
\label{fig:MI_z}
\end{center}
\end{figure}

\section{Data}\label{sec:data}

\subsection{SDSS Data}\label{sec:sdss}
We assemble our parent quasar sample from a combination of the SDSS quasar catalogues from DR7 \citep{schneider10} and DR10 \citep{paris14}. The location within the SDSS colour-space provides the basis for the selection of most quasar candidates, with \cite{richards02} describing the original SDSS selection criteria. For an object with an ultraviolet excess (typical for quasars) to be selected for spectroscopic follow-up, it should be brighter than $i \sim$19.1 ($i \sim$ 20.2) if its estimated redshift based on its colours is $z \la 3$ ($z \ga 3$). However, the quasar candidate selection criteria applied up through DR7 were biasing the selection against objects with $z \ga 2.2$. The Baryon Oscillation Spectroscopic Survey (BOSS) quasars included in DR10, on the other hand, were sought for in the redshift range between 2.15 and 3.5, down to fainter magnitudes ($g<22.0$ or $r<21.85$), in order to reach a higher quasar surface density \citep{ross12}. We include both DR7 and DR10 quasars to get a larger and more uniform redshift coverage. Had we opted for DR7 (DR10) alone, we would have missed the bulk of the $z > 2$ ($z<2$) quasars. The resulting combined sample, though not complete in redshift, covers a more representative space in absolute $i$-band magnitude, $M_\textit{i}$, and redshift than would either of the two data releases individually. 

As a function of redshift, DR7 is more than 90 per cent complete up to redshifts of about 2.2 \citep{richards02}. Thus, DR10 was designed to target objects above this redshift; that is, quasars at $2.15 \le z \le 3.5$, over which the CORE selection is homogeneous and uniform \citep{ross12, white12}. Within this range, DR10 has a given maximum completeness from single-epoch data of seventy per cent at $z=2.2$ \citep[][fig. 14]{ross12}. As such, some of the highest redshift BAL quasars might be missing from our sample.

The combined use of DR7 and DR10, without any restrictions, potentially affects the overall completeness and uniformity of the sample, since the different apparent magnitude cuts in the two releases could make the combined sample incomplete in terms of redshift. K-S tests on the distribution of the various colours and $M_\textit{i}$ however yield $p\,$-\,values (all above 0.4) that cannot reject the null hypothesis that the DR7 and DR10 sub-samples come from the same colour and $M_\textit{i}$ parent distributions. Moreover, \cite{ma15}, who also study the FIR properties of a combined sample of quasars from DR7 and DR10, similarly reach the conclusion that any incompleteness resulting from the non-uniform selection of targets in the two releases does not affect the validity of such a study.

\subsection{Herschel Data}\label{sec:sample}
We look for \textit{Herschel} counterparts to our SDSS quasar sample in the {\it Herschel} Multi-tiered Extragalactic Survey \citep[HerMES;][]{oliver12} and the {\it Herschel} Stripe 82 Survey \citep[HerS;][]{viero14}. HerS consists of 79 deg$^2$ of contiguous SPIRE imaging of the SDSS ``Stripe 82''. The HerMES fields considered here are the HerMES Large Mode Survey (HeLMS) field, the widest HerMES tier spanning 270 deg$^2$ observed with SPIRE alone, as well as the northern Level 5 and 6 fields of Bootes, EGS, ELAIS-N1, Lockman, FLS and \textit{XMM}-LSS, selected for their overlap with the SDSS footprint. The total area we select from is thus $\sim$ 350 deg$^2$. 

The point source catalogues from which we extract the SPIRE fluxes in the smaller HerMES fields are presented in \cite{wang14}. The HeLMS catalogue is part of the HerMES DR4 and is also briefly discussed in \cite{wang15}. The HeLMS SPIRE maps are also part of the HerMES DR3. Here we give a summary of the point source catalogue creation, as well as its completeness, accuracy and reliability. For a full description, see \cite{clarke16}\footnote{Doctoral thesis, available through Sussex Research Online: \href{http://sro.sussex.ac.uk/61474/}{http://sro.sussex.ac.uk/61474/}}. The HeLMS point source catalogue was created using the HerMES pipeline, a combination of the \textsc{starfinder} and \textsc{desphot} algorithms. \textsc{Starfinder} \citep{diolaiti00} is an iterative source-finding algorithm initially developed for crowded stellar fields. As such, \textsc{starfinder} can de-blend confused {\it Herschel} sources to find positions, but it requires the background to be mostly free from source flux. In {\it Herschel} SPIRE maps, however, the sky is comprised almost entirely of flux from sources. As such, \textsc{starfinder} does not accurately estimate fluxes. Thus, \textsc{desphot} \citep[De-blended SPIRE Photometry algorithm;][]{roseboom10,roseboom12} is used instead, with the \textsc{starfinder} objects as positional priors.

Both the HeLMS and HerS fields are contaminated by Galactic cirrus. This is in contrast to the other HerMES fields on which the source extraction techniques were developed, which are relatively free from cirrus; thus, both the maps and the source extraction techniques have to be modified. FIR emission from cirrus boosts fluxes of sources behind it such that a naive source extraction will find more and brighter sources in a cirrus-contaminated region. Cirrus has structure on large scales, as opposed to the small-scale fluctuations in the map from extragalactic sources, so emission from cirrus can be removed using a high-pass filter. For consistency, we adopt the filtering technique applied for HerS, as defined in \cite{viero14}, for the creation of the HeLMS catalogue. 

The confusion noise and background have been calculated from the entire map rather than on individual tiles, into which the large map has been segmented at the time of source extraction \citep[for details see][]{viero14}. The total error on each source is given as $\sigma_{\rm{T}} = \sqrt{\sigma_{\rm{conf}}^2 + \sigma_{\rm{inst}}^2}$, with $\sigma_{\rm{inst}}^2$ as the instrumental noise. The confusion noise, $\sigma_{\rm{conf}}^2$, has been calculated on the residual map by fitting a line to the flux as a function of coverage, where the confusion noise is the y-intercept. Any source with $\chi^2 > 10$ is removed from the catalogue, leaving a total of 92,256 sources at 250\,\micron.

To test the positional and flux accuracy, we inject sources of known fluxes at random positions in the map. The maps are then run through the source extraction pipeline again. We find the positional accuracy to be within 5 arcsec in both right ascension and declination. The flux accuracy is already above 90 per cent at 30\,mJy and reaches 98 per cent at 40\,mJy. 

Finally, to match the catalogues in the three SPIRE bands and assess completeness, we follow \cite{smith12}. We find a maximum completeness of 93 per cent, attained at $\sim$60\,mJy. The remaining sources are missed by the \textsc{starfinder} algorithm, but are not associated with any particular region of the map \citep[see also][]{wang14}.

\subsection{SDSS-\textbf{\textit{Herschel}} matching}\label{sec:matching}
\cite{wang14} have shown through simulations, in the fashion mentioned above, that the positional accuracy of sources in the DR2 HerMES point source catalogues also peaks at 5 arcsec. We therefore look for SDSS DR7 and DR10 quasars with $>3\sigma$ detections at 250\,$\micron$ applied across all the fields, using a 5 arcsec matching radius between the SDSS and {\it Herschel} catalogues. The results of this match are provided in Table \ref{tab:jointDetections}. Note that we rely on the individually detected objects only and do not stack the SPIRE fluxes of the non-detected SDSS quasars. Our final catalogue contains 83, 212 and 218 unique SDSS quasars (i.e. they only appear once in the catalogue even if they belong to both DR7 and DR10) in the HerMES fields, HerS and HeLMS, respectively. In what follows, we call these 513 quasars the {\it Herschel}/SDSS quasar sample. 

The {\it Herschel}/SDSS quasars span the redshift range $0.1627 \le z \le 4.679$, and the bulk lie in the $M_\textit{i}$ \citep[as defined in][]{schneider10} range between -22.5 and -29. The $M_\textit{i}-z$ plane of the sample is shown in Fig.  \ref{fig:MI_z}, where the DR7 and DR10 objects are marked in red circles and blue squares, respectively. The filled symbols denote BAL quasars and will be discussed separately. The histograms of the quasars belonging to the DR7 and DR10 releases, as well as that of the final sample, are shown in the top figure as filled red bars, dashed blue and solid black lines, respectively. The green spikes show the redshift distribution of the BAL quasars.

\begin{figure}
\begin{center}
\vspace*{-2cm}
\includegraphics[width=8cm]{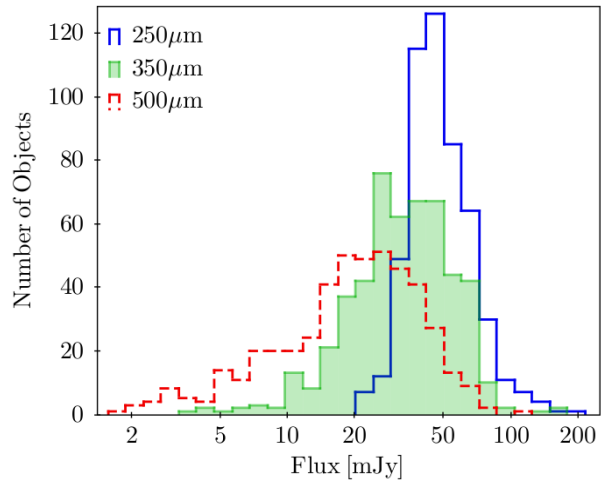}
\vspace*{-1.5cm}
\caption{SPIRE flux distributions of the {\itshape Herschel}/SDSS quasar sample. In the solid blue, filled green and dashed red histograms, we show the 250, 350 and 500\,$\micron$ fluxes, respectively.}
\label{fig:flux250}
\end{center}
\end{figure}

The 350 and 500\,$\micron$ fluxes are extracted on the 250\,$\micron$ positions, but we apply no cuts in signal-to-noise for either the 350 or the 500\,$\micron$ fluxes. This results in 505 detections at 350\,$\micron$ and 473 detections at 500\,$\micron$. The SPIRE flux distributions of the {\it Herschel}/SDSS quasar sample are shown in Fig. \ref{fig:flux250}. 

The \textit{Herschel} detection rates at 250\,$\micron$ of SDSS quasars in our fields average at $\sim$5 per cent, as seen when comparing the 3rd and 6th columns of Table \ref{tab:jointDetections}, with a constant detection fraction up to $M_i \sim$ -27 and an increase by a factor of $\sim$1.5 towards brighter absolute magnitudes. This number is smaller than the detection rates mentioned in previous works (e.g. \citealt{hatzimi10}, who find a 30 per cent detection rate in Lockman and FLS at 5$\sigma$ but for both type 1 and type 2 AGN combined). The lower detection rates within our sample are largely due to the fact that HerS and HeLMS, where 85 per cent our quasars are located, are shallower than are the HerMES fields. Our detection rates are instead more in line with the 8 per cent detection rate presented in \cite{bonfield11}, using a simple positional cross-match in the H-ATLAS science demonstration field. 

\begin{figure*}
\begin{center}
\includegraphics[width=8cm]{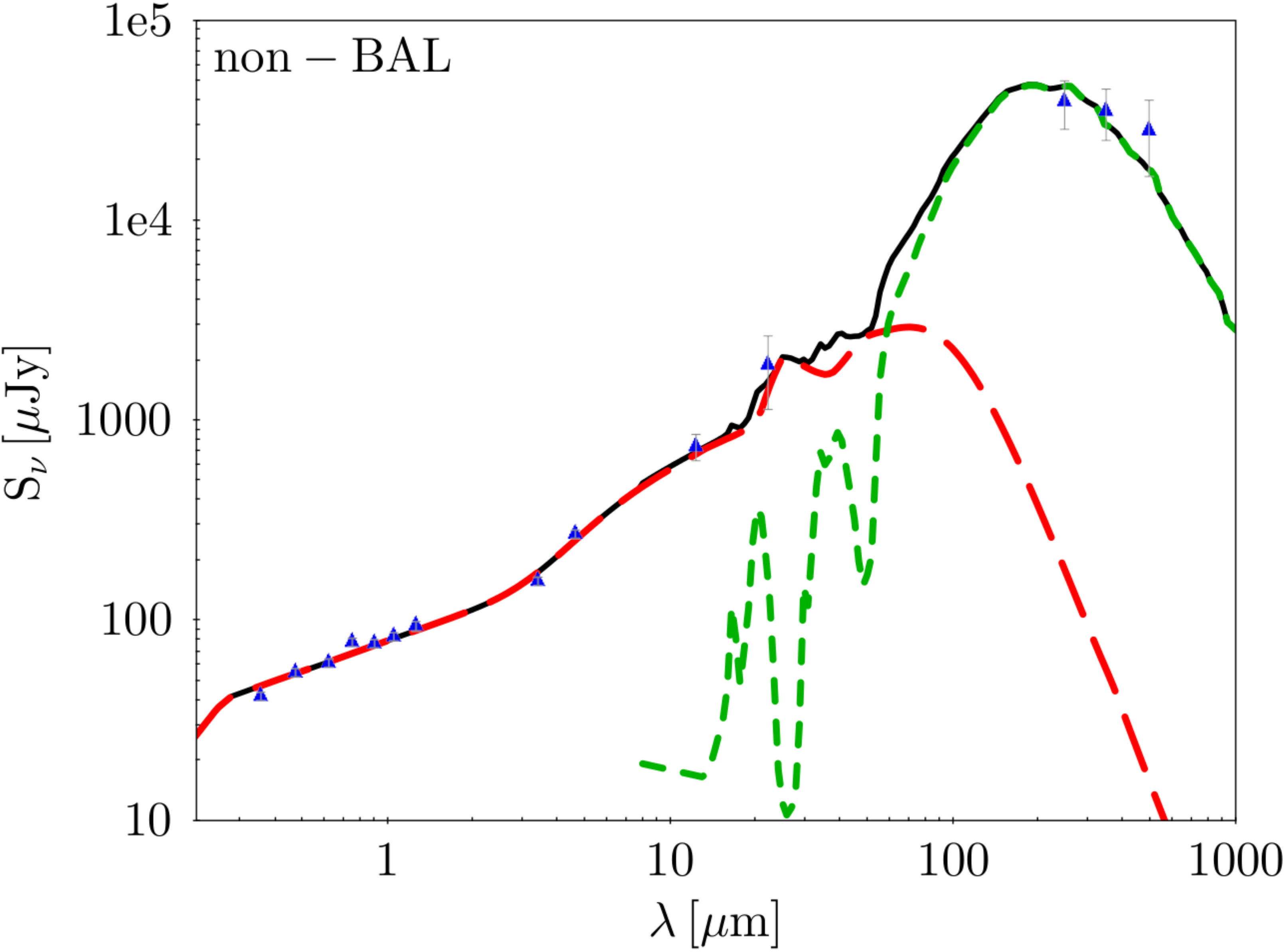}
\hspace{0.5cm}
\includegraphics[width=8cm]{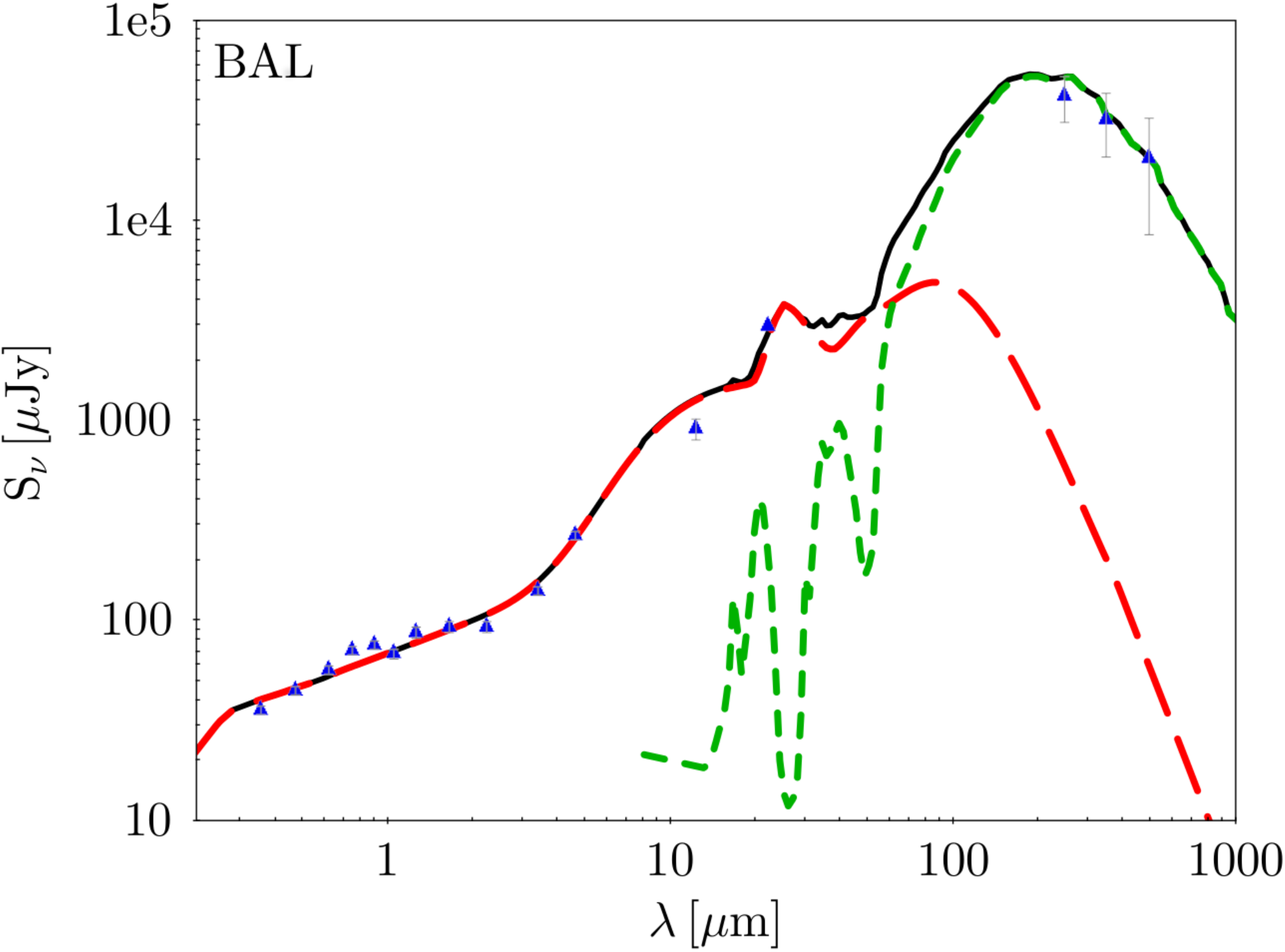}
\caption{Example SED fits. The left panel shows the fit for one of our general population quasars, while the right panel shows a fit for one of our BAL quasars. The photometric data are shown as blue triangles. The total fits are shown in solid black, while the two components of the fit are shown in the long dashed red (AGN) and the short dashed green (starburst) lines.}
\label{fig:sed}
\end{center}
\end{figure*}

\subsection{Ancillary data}\label{AncillaryData}
Of our 513 quasars, 473 are also detected by the Wide-Field Infrared Survey Explorer \citep[WISE;][]{wright10}. We also assemble near-infrared detections from three surveys: the Two Micron All Sky Survey \citep[2MASS;][]{skrutskie06}, the UKIRT Infrared Deep Sky Survey \cite[UKIDSS;][]{lawrence07}, and the VISTA Hemisphere Survey \citep[VHS;][]{mcmahon13} Data Release 3. There are 196 objects in the HerS catalogue with a measurement in one or more of the UKIDSS bands and 178 within HeLMS. There are also 64 objects (16, 14 and 34 in HerMES, HerS and HeLMS, respectively) with VHS photometry. Finally, 29 objects have a measurement in at least one of the 2MASS bands. While HerS and HeLMS also have 2MASS data available, we choose to use the UKIDSS data, as it is deeper. As a result, for more than 90 per cent of our objects, there is a minimum of 11 and a maximum of 16 photometric data points available from the optical to the FIR.

\section{Analysis}\label{sec:analysis}

\subsection{Starburst and AGN luminosities}\label{sec:sedfit}
To infer starburst and AGN luminosities, we use the multi-component SED fitting technique described in \citealt{hatzimi08,hatzimi09}. In the most general case, the method fits fluxes and their errors with three separate models: stellar, AGN, and starburst. Our sample however consists entirely of luminous type 1 quasars with extremely high SFRs. This allows us to make two key simplifications to the general method. First, it is almost certain that the stellar emission will be outshone by the AGN and/or the starburst at all wavelengths, so we do not include a stellar template in the fit. Second, since our sources are type 1 quasars, it is reasonable to conclude that the optical to mid-infrared (MIR) coverage will be dominated by the AGN \citep{hatzimi05}, and that the FIR coverage will be dominated by the starburst \citep[e.g][]{hatzimi10,ma15,harris16}. The infrared emission reprocessed by the AGN torus strongly depends on the dust geometry. In particular, the ratio between the outer and inner radius of the torus plays a major role in determining the AGN SED at longer wavelengths (in the models used in this paper the maximum value of this ratio is 150). As a result, the peak of the infrared emission of the AGN torus models lies at $10-30$\,$\micron$ (fig. 5 of \citealt{feltre12}) and the maximum width of the infrared bump (as defined in \citealt{feltre12}) for these models is $\sim 50$\,$\micron$. We therefore do not expect any significant contribution from the AGN emission at wavelengths larger than $\sim 50$\,$\micron$. Complementary, observationally motivated discussion on this point can be found in \S$5.3$ of \citealt{harris16}. We note the caveat though that, with our data, which do not cover the range between 22 and 250\,\micron, we cannot completely rule out significant FIR emission from the AGN. We thus use the shorter wavelength data to constrain the properties of the AGN, and, having done so, then use the SPIRE data to constrain the properties of the starburst. Two example SED fits are given in Fig. \ref{fig:sed}; one for a classical quasar and one for a BAL quasar.

The AGN component of the fit includes both the direct emission and the reprocessed emission, that is, optical/ultraviolet light that is absorbed by the dust in the torus and reradiated in the infrared. We call the luminosity of the best-fitting AGN model the accretion luminosity, $L_\textnormal{acc}$. We have assumed the smooth, continuous distribution of dust within the torus described in \cite{fritz06} and \cite{feltre12}. This set of models has been shown to accurately reproduce the SEDs of both obscured and unobscured AGN \citep[e.g.][]{fritz06, hatzimi08, hatzimi09, gruppioni16}. We constrained the fit to a subset of 216 torus templates (out of the $\sim$3700 models in the full grid), selecting a representative set of parameters defined in \cite{hatzimi08}. This is done to minimise degeneracies among model parameters that we do not seek to constrain, without affecting the quality of the fits for the luminosities, given that there is a large overlap in SED shapes when varying the model parameters \citep[for a description of the effect of each of the parameters on the SED, see][]{fritz06}.

The starburst component is modelled with a library of templates of six starburst galaxies: Arp220 (merger), M82 (disk/amorphous), M83 (SAB(s)c), NGC1482 (E/int), NGC4102 (Sb), and NGC7714 (Sc).  The starburst templates span systems with low (e.g. M82 \& M83, \citealt{nikola12,foyle13}) to high (e.g. Arp 220, \citealt{farrah03,rang11}) extinctions and a broad range in mid- to far-IR SED shapes, star-formation histories, and Hubble types. Thus, they plausibly span at least a substantial fraction of the properties of star-forming galaxies at $z>0.5$ and have been used to model starbursts at all redshifts before \citep[see e.g.][]{feltre13}. The luminosity of the starburst component, $L_\textnormal{SB}$, is that of the best-fitting template scaled to the observed data points, integrated between 8 and 1000\,$\micron$. Finally, $L_\textnormal{SB}$ is then converted into an SFR using the original (i.e. assuming a Salpeter IMF) relation derived by \cite{kennicutt98}

\begin{equation}\label{eq:kenn}
\frac{\textnormal{SFR}}{\textnormal{M}_\odot \hspace{0.1cm} \textnormal{yr}^{-1}} = 4.5 \times 10^{-44} \frac{L_\textnormal{SB}}{\textnormal{erg} \hspace{0.1cm} \textnormal{s}^{-1}} 
\end{equation}

\begin{figure}
\begin{center}
\vspace*{-2cm}
\includegraphics[width=7.5cm]{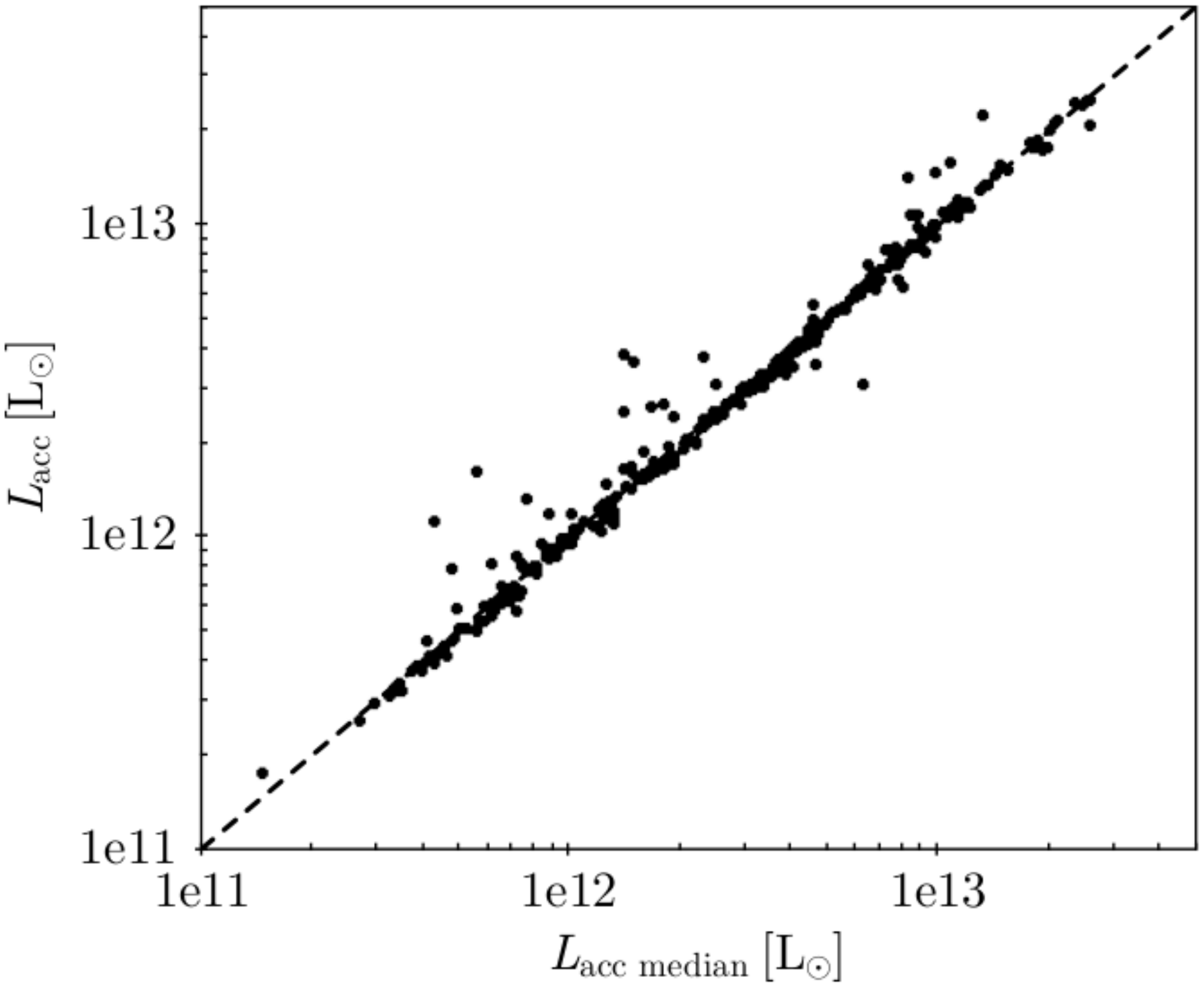}\\
\vspace*{-3cm}
\includegraphics[width=7.5cm]{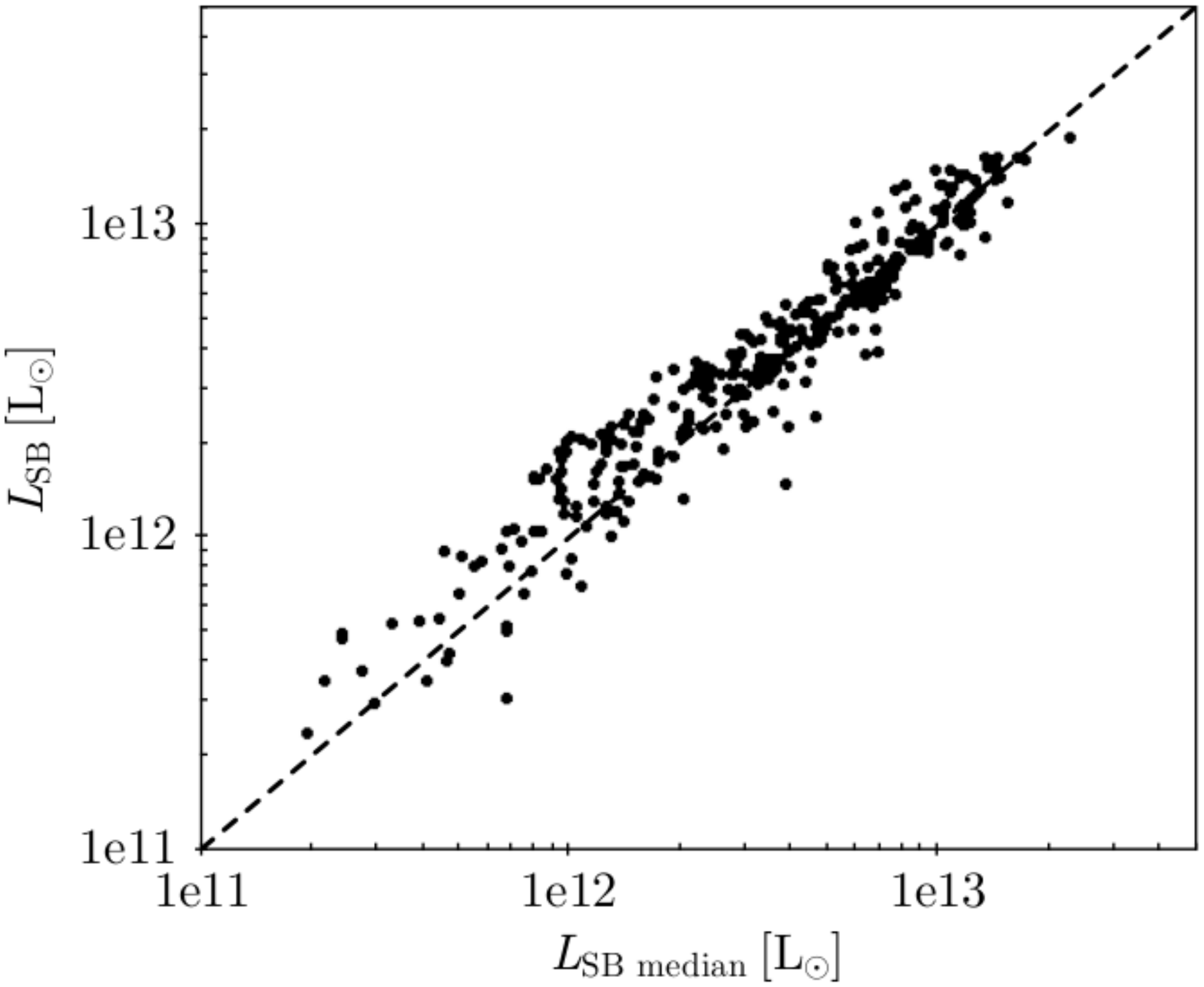}\\
\vspace*{-1.5cm}
\caption{A comparison between the luminosities extracted from the best-fitting SEDs and from the median of their respective PDFs for $L_\textnormal{acc}$ (top) and $L_\textnormal{SB}$ (bottom). The agreement is excellent in both cases.}
\label{fig:lcomp}
\end{center}
\end{figure}

The $L_\textnormal{acc}$ and $L_\textnormal{SB}$ values derived from this approach have been shown to be robust \citep[see][]{hatzimi08}. To estimate the uncertainties on $L_\textnormal{acc}$ and $L_\textnormal{SB}$ specific to our set of templates, we calculate the median of the probability distribution function (PDF) of the values obtained using all combinations of models in the fit. Fig. \ref{fig:lcomp} compares the best fit $L_\textnormal{acc}$ (top) and $L_\textnormal{SB}$ (bottom) with the median of their respective PDFs. The root-mean-square deviation of the two sets of distributions are 7 per cent and 15 per cent, respectively. In other words, the best-fitting $L_\textnormal{acc}$ ($L_\textnormal{SB}$) differs, on average by 7 (15) per cent from the median value of the PDF.

\subsection{Black hole masses \& Eddington ratios}\label{sec:agn_prop}
We estimate black hole masses ($M_\textnormal{BH}$) from the FWHM of the C\begin{scriptsize}IV\end{scriptsize}, Mg\begin{scriptsize}II\end{scriptsize} or H$\beta$ lines (e.g. \citealt{wandel99, kaspi00, mclure02}), using: 

\begin{equation}\label{eq:Mbh_form}
\frac{M_\textnormal{BH}}{\textnormal{M}_{\odot}} = \alpha \left(\frac{\lambda L_{\lambda}}{10^{37} \textnormal{W}} \right)^{\beta} \left( \frac{\textnormal{FWHM}}{\textnormal{km} \hspace{0.05cm} \textnormal{s}^{-1}} \right)^2
\end{equation}

\noindent with $\alpha$\,\,=\,\,4.6, 5.2 and 8.2, $\beta$\,\,=\,\,0.53, 0.62 and 0.50, and $\lambda$\,\,=\,\,1350 \AA, 3000 \AA \, and 5100 \AA \, for C\begin{scriptsize}IV\end{scriptsize}, Mg\begin{scriptsize}II\end{scriptsize} and H$\beta$, respectively, as described in \cite{vester06} and \cite{shen11}, the latter of which offers a recalibration of the Mg\begin{scriptsize}II\end{scriptsize} formula initially described by \cite{mclure04}. We then estimate Eddington ratio as $\lambda_{Edd} = L_\mathrm{acc}/{L_\mathrm{Edd}}$, where $L_\mathrm{Edd} $ is the Eddington luminosity, calculated as 

\begin{equation}\label{eqnLed}
\frac{L_\mathrm{Edd}}{\rm{erg}\,\rm{s}^{-1}} = \frac{4 \pi  G m_{p} c} {\sigma_{T}} {\it M}_{\rm{BH}} = 1.26 \times 10^{38}\frac{{\it M}_{\rm{BH}}}{\rm{M}_{\odot}}
\end{equation}

\noindent in which $\sigma_{T}$ is the Thompson scattering cross-section for the electron, $m_{p}$ is the mass of the proton, $c$ is the speed of light and $G$ the gravitational constant.

The uncertainties on the measurements of the FWHMs of C\begin{scriptsize}IV\end{scriptsize}, Mg\begin{scriptsize}II\end{scriptsize}, and H$\beta$ are of the order of 10 per cent \citep{shen11}. The same authors discuss in detail the systematic errors on $M_\textnormal{BH}$ introduced by the use of the different emission lines. They show the offset between the estimates using Mg\begin{scriptsize}II\end{scriptsize} and C\begin{scriptsize}{IV}\end{scriptsize} to be negligible. For those objects in the redshift range such that both the C\begin{scriptsize}{IV}\end{scriptsize} and Mg\begin{scriptsize}{II}\end{scriptsize} lines are present in the spectra, we use the Mg\begin{scriptsize}{II}\end{scriptsize} line as the BH mass tracer. There are no relevant emission line data available to compute the BH masses for eleven of our objects; we retain these objects in the sample but discard them from the relevant sections.

\subsection{Complementarity with other studies}\label{sec:comple}
This paper complements and builds upon other works on the star-forming properties of quasar host galaxies. We here describe how it relates to the two most closely associated works, those of \citealt{ma15} and \citealt{harris16}, hereafter MY15 and H16, respectively.

In many ways, the present work is an expansion of the work in MY15, who also study the star-forming properties of {\it Herschel}-selected SDSS quasars, using a sample of 354 quasars in all the HerMES DR2 catalogues, regardless of the depth of the fields, as well as the HerS and H-ATLAS \citep{eales10} catalogues. Our sample, on the other hand, includes the HeLMS data, never previously been used in a study of this type, as well as HerS and the wider (and shallower) northern HerMES fields listed in Sec. \ref{sec:sample}. Moreover, while MY15 match SDSS positions to the positions in the HerMES, HerS and H-ATLAS public catalogues using a matching radius of 3 arcsec, we adopted a less conservative matching radius of 5 arcsec based on the simulations in \cite{wang14}. From a visual inspection of all the sources, we find the resulting matches to be highly reliable. As a result of all the above, our {\it Herschel}/SDSS sample has 187 objects (about a third of its size) in common with the sample in MY15. 

Our approach to extracting SFRs also differs to that used by MY15. MY15 considered  the coldest dust component alone, by fitting a single-temperature modified black body to the SPIRE data.  For testing purposes, a two-component SED fitting including an AGN and a starburst component was performed on a handful of objects with available PACS photometry. Conversely, in this work we constrain the AGN properties by focusing on the AGN-dominated part of the SED, using data in the near-IR (2MASS, UKIDSS or VHS photometry) and MIR (WISE), and by using a set of AGN models that have proved to accurately reproduce the optical-to-MIR quasar SED. Therefore, the key difference in the derived SFRs is that, in this work, the SFRs are measured on the full range 8-1000\,$\micron$, including the contribution to the MIR.

Turning to H16, our paper complements and extends this work. In H16, the authors also use {\it Herschel} data to study the star-forming properties of SDSS quasars. However, H16 focus on quasars in a narrower, systematically higher redshift range, $2<z<3$, and have a sample which is mostly individually {\itshape un}detected by {\it Herschel}. Hence their analysis relies on stacked observations rather than individual detections. At the same redshift, therefore, we study quasars with much higher SFRs on average.

\section{Results}\label{sec:results_full}

\subsection{Star formation rates}\label{sec:sfrs_full}
\begin{figure}
\begin{center}
\includegraphics[width=7.5cm]{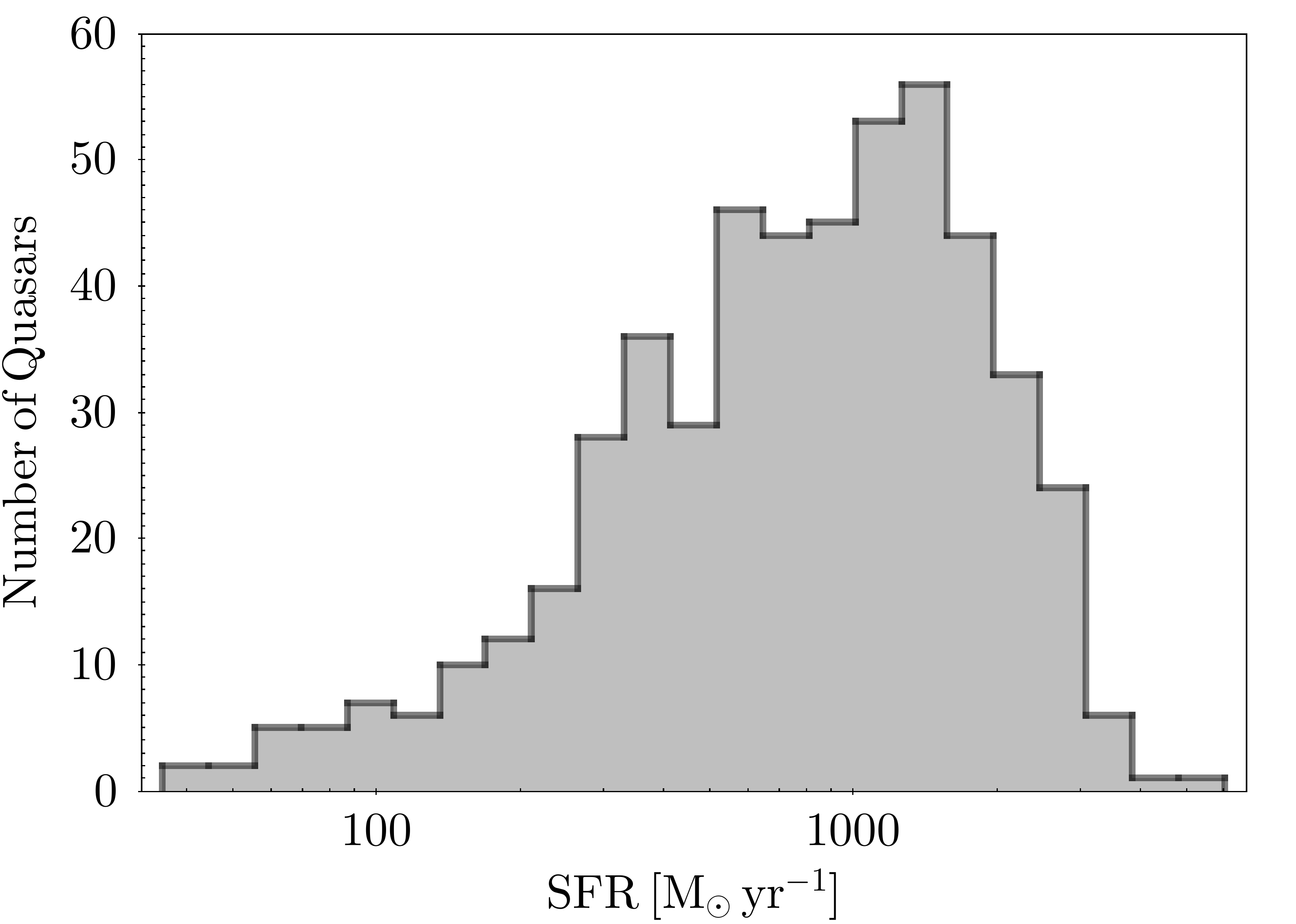}
\caption{The SFR distribution for the {\it Herschel}/SDSS quasar sample. Our sample represents some of the most extreme starburst systems, as evidenced by the high inferred SFRs, which places them above the main sequence, for their respective redshifts, by over a dex on average.}
\label{fig:sfr_full}
\end{center}
\end{figure}

The distribution of the SFRs in our sample is shown in Fig. \ref{fig:sfr_full}. The SFRs are extremely high, with a mean of $1000\substack{+1400 \\ -600}$\,M$_\odot \hspace{0.1cm} \textnormal{yr}^{-1}$. This aligns well with those rates estimated for other individually FIR-detected populations, such as submillimetre galaxies (e.g. \citealt{chapman10,wardlow11,barger14}) and infrared-bright radio galaxies (e.g. \citealt{drouart14}), over a similar redshift range. The mean SFR is a factor $\sim$2 higher than that in MY15\footnote{MY15 quote 50 per cent of their objects exhibiting SFRs above 300\,M$_\odot \hspace{0.1cm} \textnormal{yr}^{-1}$, derived using Chabrier IMF. The value would be $\sim$510\,M$_\odot \hspace{0.1cm} \textnormal{yr}^{-1}$ assuming a Salpeter IMF.}. The difference is likely due to the fact that MY15 only considered SFRs measured from the coldest components, while our SFRs also include emission in the MIR.

Next, we examine the behaviour of the SFR as a function of $L_\textnormal{acc}$ (as defined in Sec. \ref{sec:sedfit}), $M_\textnormal{BH}$, and $\lambda_\textnormal{Edd}$. To account for the fact that this is a flux-limited sample, we split the sample into three redshift bins, $z<1.0$, $1.0 \le z < 2.0$ and $z \ge 2.0$. Each redshift bin is further split into $L_\textnormal{acc}$, $M_\textnormal{BH}$, or $\lambda_\textnormal{Edd}$ bins, and the average SFR is computed in each of these bins. 

\begin{figure}
	\begin{center}
		\includegraphics[width=9cm]{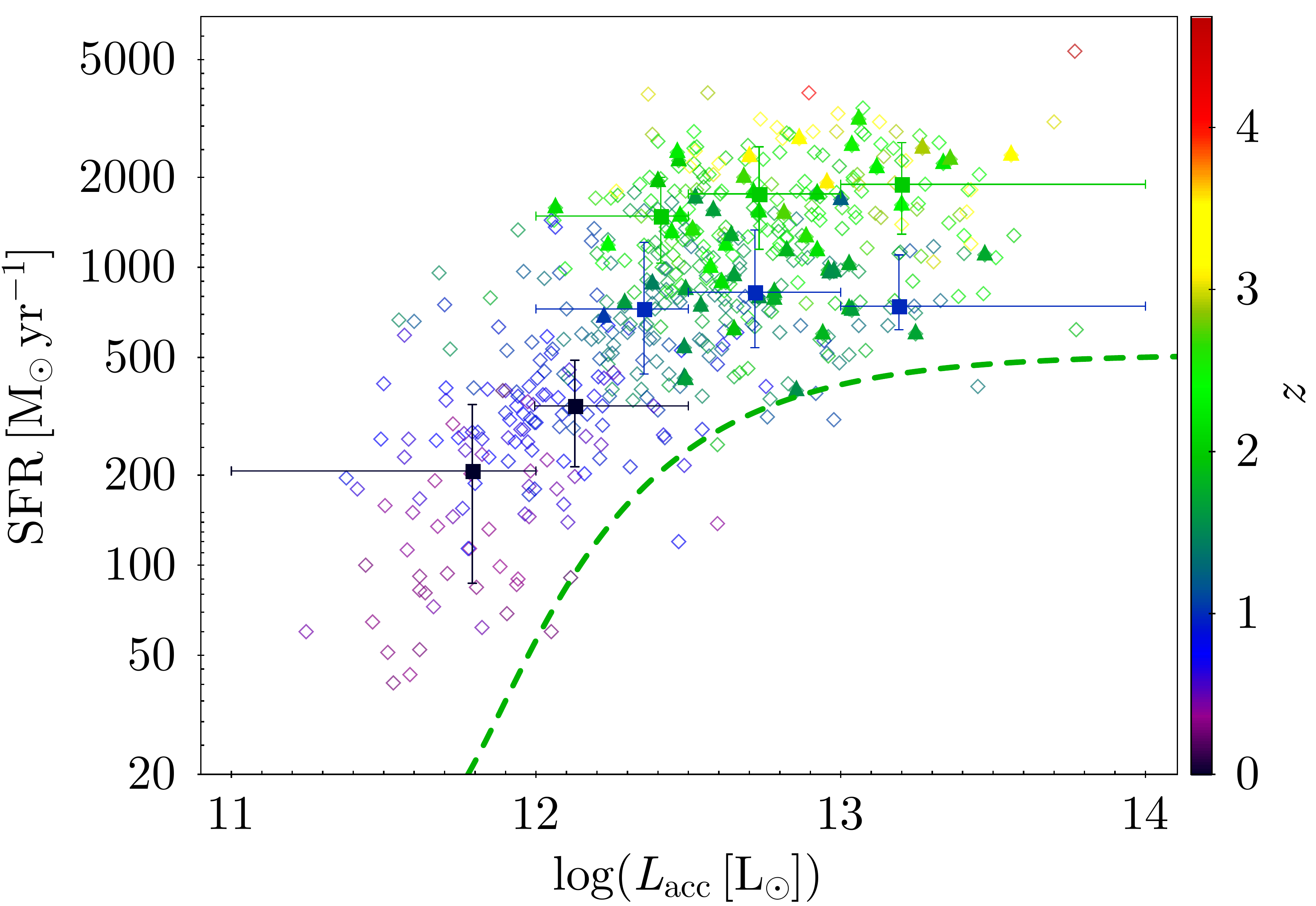} \\
		\vspace*{.15cm}
		\includegraphics[width=9cm]{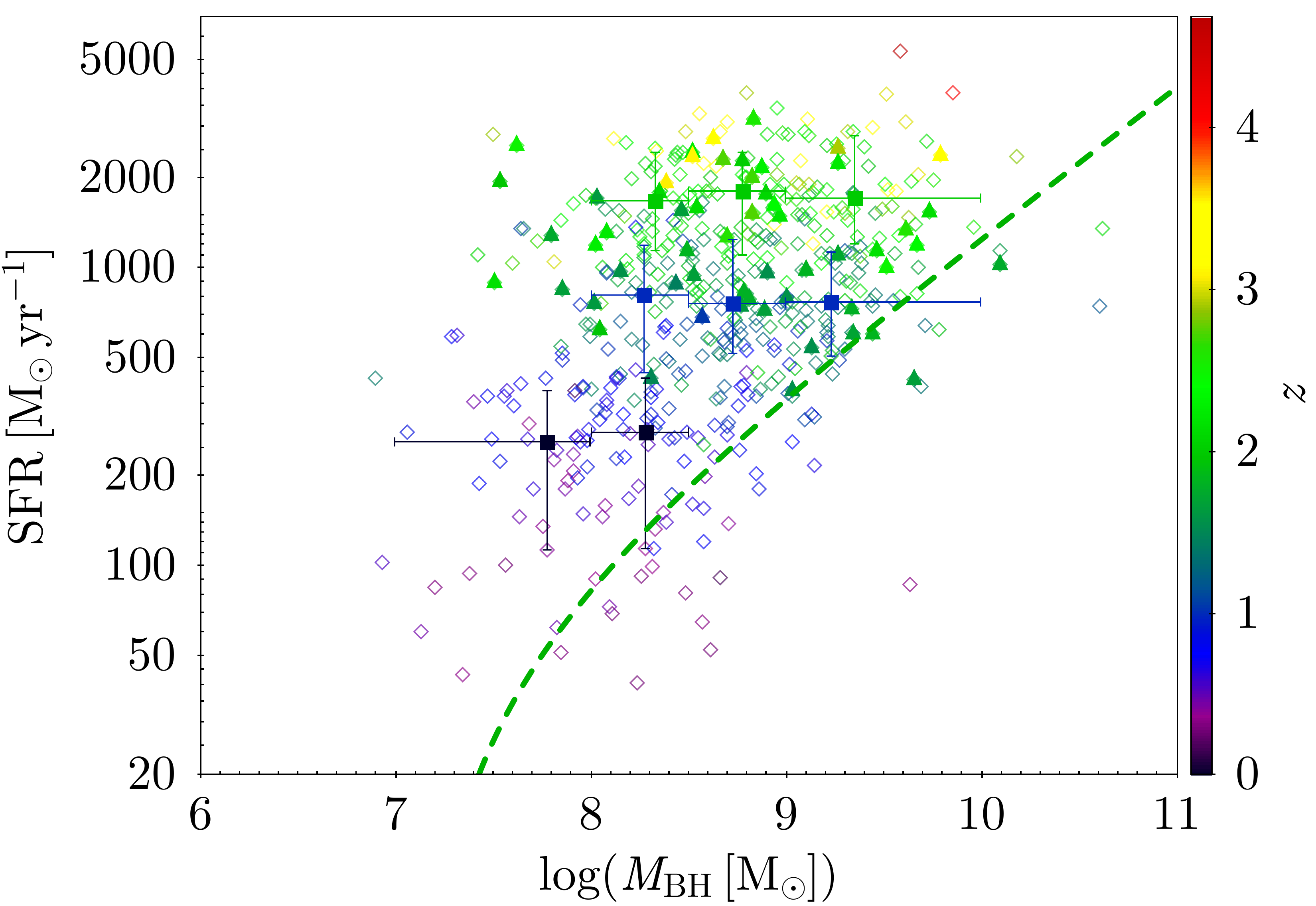} \\
		\vspace*{0.15cm}
		\includegraphics[width=9cm]{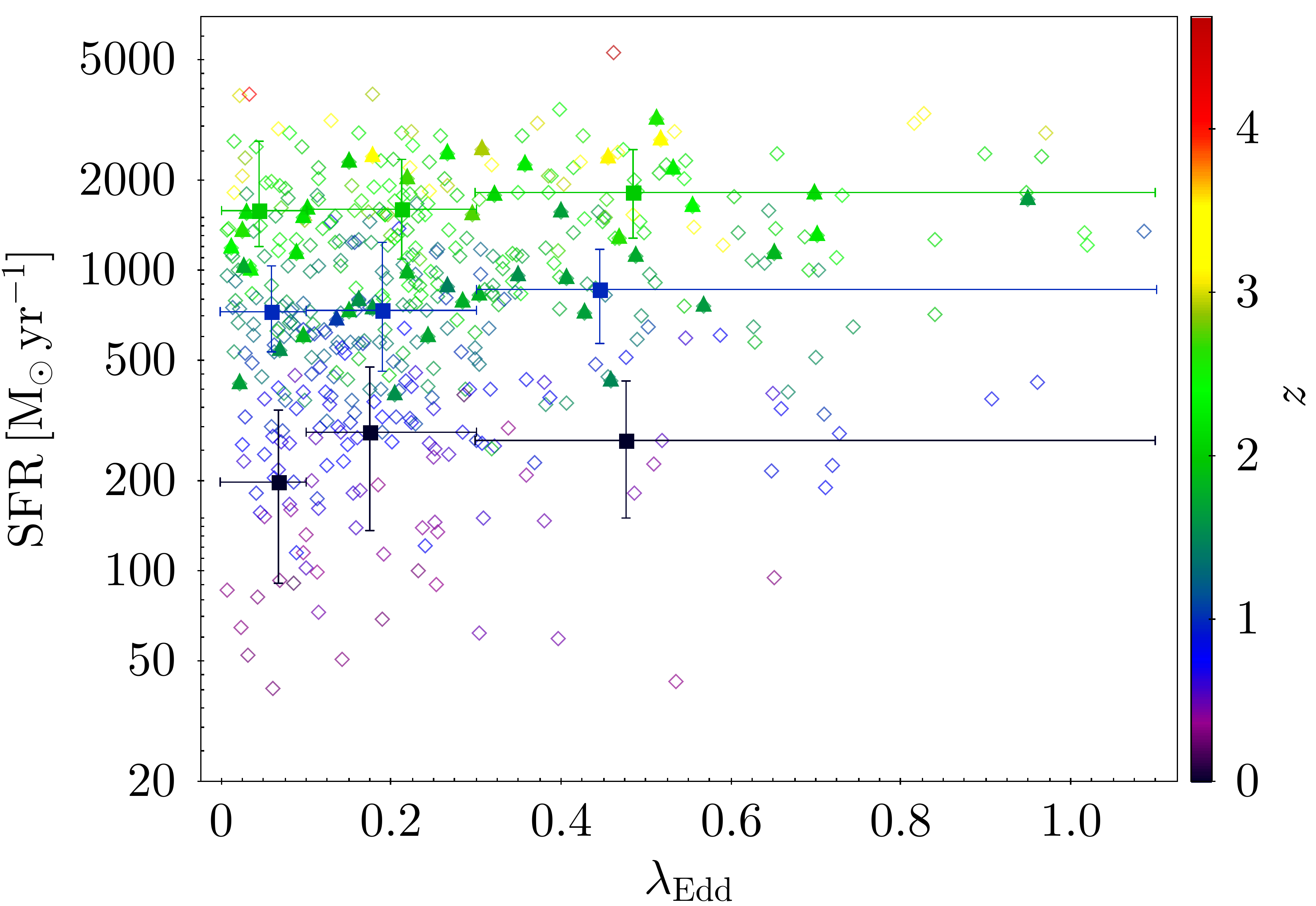}
		\caption{SFR versus $L_\textnormal{acc}$ (top panel), $M_\textnormal{BH}$ (middle panel) and  $\lambda_\textnormal{Edd}$ (bottom panel), colour-coded by redshift. The three redshift bins are $z<1.0$, $1.0 \le z < 2.0$ and $z \ge 2.0$. SFRs increase with redshift, but do not correlate with any of the intrinsic quasar properties within the same redshift range. The green lines in the top and middle panels show the relations described in \protect\cite{harris16} for quasars at $2 < z< 3$ with a factor $2-3$ lower SFRs than in our sample. The filled squares and associated error bars show the mean value in each bin. The diamonds and triangles show the results for the non-BAL and BAL subsamples, respectively.}
		\label{fig:SFR_trends}
	\end{center}
\end{figure}

We find the SFR to increase by a factor of $\sim$8 from the low ($z<1.0$) to the high ($z \ge 2.0$) redshift bins for quasars of the same $L_\textnormal{acc}$. In a given redshift interval, however, the SFR is consistent with being constant for increasing $L_\textnormal{acc}$ for all redshift bins. This is shown in the top panel of Fig. \ref{fig:SFR_trends}, where the average points are overplotted on the individual objects. 
The typical uncertainties on $L_\textnormal{acc}$ and $L_\textnormal{SB}$, discussed in Sec. \ref{sec:sedfit}, are small relative to the size of the bins and have no effect on the results. The green line shows the relation described in H16 and will be discussed in Sec. \ref{sec:discussion}. The bins in Fig. \ref{fig:SFR_trends}, however, are large, such that the apparent lack of a trend might be a result of degeneracies between redshift and luminosity. We tried a finer binning, with bins half the size of the ones presented here in terms of both redshift and luminosity, and found consistent results. We also found consistent results by examining SDSS DR7 and DR10 separately. The (lack of) trends are, therefore, not a bias as far as we can tell.

The middle panel of Fig. \ref{fig:SFR_trends} plots SFR against $M_\textnormal{BH}$. We again see that the SFR is constant with increasing $M_\textnormal{BH}$ inside a given redshift bin. Here there is no hint of a rise in SFR; while the uncertainties on the SFR and $M_\textnormal{BH}$ are significant, the data are entirely consistent with a flat relation. Finally, in the bottom panel of Fig. \ref{fig:SFR_trends}, we plot SFR against the Eddington ratios of the quasars. We again see no relation.

\subsection{Broad absorption line quasars}\label{sec:BAL}
A quasar is classified as BAL quasar if the equivalent width of at least one absorption line (usually C\begin{scriptsize}{IV}\end{scriptsize}) exceeds 2000\,$\textnormal{km} \hspace{0.1cm} \textnormal{s}^{-1}$ \citep{weymann91}. BAL quasars can be further split into three sub-populations. The most numerous of these is that of the high-ionisation BAL (HiBAL) quasars. HiBAL quasars display absorption in Ly$\alpha \hspace{0.025cm} \lambda$1216\textup{\AA}, N\begin{scriptsize}{V}\end{scriptsize}$\hspace{0.025cm} \lambda$1240\textup{\AA}, Si\begin{scriptsize}{IV}\end{scriptsize}$\hspace{0.025cm} \lambda$1394\textup{\AA} and C\begin{scriptsize}{IV}\end{scriptsize}$\hspace{0.025cm} \lambda$1549\textup{\AA}. In addition to the absorption features found in HiBAL quasars' spectra, low-ionisation BAL (LoBAL) quasars also display absorption features in such low-ionisation ions as Al\begin{scriptsize}{III}\end{scriptsize}$\hspace{0.025cm} \lambda$1857\textup{\AA} and Mg\begin{scriptsize}{II}\end{scriptsize}$\hspace{0.025cm} \lambda$2799\textup{\AA}. The third class, FeLoBAL quasars, shares the same properties as LoBAL quasars with additional absorption features from excited levels of iron.

\begin{table}
	\begin{center}
		\begin{tabular}{L{1.15cm} C{2cm} C{2.05cm}}
			\hline
			Field & SDSS \hspace{3cm} BAL quasars & \textit{Herschel}/SDSS BAL quasars \\
			\hline
			\noalign{\vskip0.1cm}
			HerMES & 92  & 7  \\
			HerS   & 344 & 20 \\
			HeLMS  & 666 & 32 \\
			\hline
		\end{tabular}
	\end{center}
	\caption{BAL detection statistics by \textit{Herschel} field. The average {\it Herschel} detection rate of BAL quasars is about 5 per cent, the same as for non-BAL quasars, as shown in Table \ref{tab:jointDetections}.}
\label{tab:bal_stats}
\end{table}

We select BAL quasars using the SDSS \texttt{BAL\char`_FLAG}, which is set to 1 when C\begin{scriptsize}{IV}\end{scriptsize} has an absorption feature, 2 when Mg\begin{scriptsize}{II}\end{scriptsize} has one and 3 when both C\begin{scriptsize}{IV}\end{scriptsize} and Mg\begin{scriptsize}{II}\end{scriptsize} show absorption. Table \ref{tab:bal_stats} shows the number of BAL quasars in each of the three \textit{Herschel} surveys with at least a $3\sigma$ detection at 250\,$\micron$. The {\it Herschel} detection rates of BAL quasars are around 5 per cent, identical to those of non-BAL quasars. Of the 59 BAL quasars detected in the three {\it Herschel} fields, 57 are HiBAL quasars and two are LoBAL quasars. As HiBAL and LoBAL quasars may be physically different \citep[see e.g.][]{streb10}, we restrict our analysis to the 57 HiBAL quasars. 

We create two sub-samples from the sources in the rectangle in Fig. \ref{fig:MI_z}. The lower redshift cut ($z$=1.5) corresponds approximately to the wavelength at which the C\begin{scriptsize}{IV}\end{scriptsize} line shifts out of the SDSS bandpass, eliminating two of the BAL quasars, while the upper redshift cut ($z$=3.3) is set where the quasar density drops significantly, and only leaves out one BAL quasar from the full set. The corresponding bright and faint absolute magnitude cuts are -28.5 and -24.3, respectively, leaving another three BAL quasars out. A total of 51 BAL and 208 non-BAL quasars reside in the defined area.  

Looking at Fig. \ref{fig:SFR_trends}, there is no preferred location for the BAL quasars in any of the parameter spaces considered (the lack of BAL quasars with low SFR is due to the lower redshift cut). To further test if the behaviour of these BAL quasars is identical to that of the non-BAL quasars, we perform K-S tests to compare the $L_\textnormal{acc}$, $\lambda_\textnormal{Edd}$ and SFR distributions of the two subsets. The BAL quasars have indistinguishable $L_\textnormal{acc}$ and $\lambda_\textnormal{Edd}$ distribution from the non-BAL quasars, as indicated by the $p$\,-\,values from the K-S test (0.47 and 0.46, respectively). Lastly, their SFR distributions are also indistinguishable ($p$\,-\,value = 0.46). We thus find no evidence that the accretion rate, accretion efficiency or star forming properties of HiBAL quasars are different from those of classical quasars.

\section{Discussion \& Conclusions}\label{sec:discussion}
The relationship between SFRs and AGN properties gives insights into the co-evolution of these activities. While we do not have stellar masses for our sample, their extremely high SFRs coupled with reasonable assumptions for quasar host stellar masses at this epoch mean they almost certainly lie above the `main sequence' for star formation at their respective epochs. In the following, we consider the relationship between SFRs and both redshift and AGN properties. Finally, we discuss the implications for BAL quasars.

\subsection{Evolution with redshift}\label{sec:disczevo}
We find that the SFRs of our sample increase by a factor of up to $\sim$\,8 from $z<1$ to $z \ge 2$. We do not see any evidence for a rise in SFRs with redshift beyond $z\sim2$. 
This is consistent with observations of the global comoving SFR density, which increases by a factor of at least ten over $0<z<1$ and peaks around $z=2$ \citep[e.g.][]{reddy08, magn09, karim11, wuyts11, bethermin13, burgarella13, wang13, chen16}. Thus, in terms of evolution over $0\lesssim z \lesssim 3$, extremely luminous star formation events in quasar hosts appear to evolve, on average, in a manner consistent with the global star formation history as derived from the general galaxy population.

Our results are also consistent with those of MY15, who find that the fraction of IR-luminous quasars peaks at $z\sim2$, and with H16, who find that SFRs in quasars do not change appreciably as a function of redshift over $2<z<3$ (but see also e.g. \citealt{netzer16}). They are also consistent with several previous studies of the evolution with redshift of star formation in AGN hosts \citep[e.g.][]{serjeant09,serjeant10,shao10,bonfield11,mullaney12}. These previous studies mostly sample different parts of the  $z$-SFR-$L_\textnormal{acc}$ parameter space for quasars to our study. For example, the sample in \citealt{mullaney12} spans a similar redshift range but has lower $L_\textnormal{acc}$ values and SFRs by $\gtrsim 2$ orders of magnitude, while the sample in H16 has comparable $L_\textnormal{acc}$ values, but lies at higher redshifts with a factor $\sim 3$ lower SFRs. The apparent consistency of our results with theirs on the evolution of SFRs with redshift in quasar hosts is thus remarkable. This consistency suggests that the physical processes that lead to star formation in AGN hosts, such as mergers or secular evolution, do so in a way that gives a similar evolution with redshift for star formation events spanning tens to thousands of $M_{\odot}$ per year and accretion luminosities $\gtrsim10^{10}$\,L$_\odot$. We caution however that this does not account for selection effects between studies, and only applies to time-scales significantly longer than the AGN and starburst duty cycles.

\subsection{Star formation and accretion luminosity}\label{sec:discsfagn}
We first consider the relation between SFR and $L_\textnormal{acc}$. The top panel of Fig. \ref{fig:SFR_trends} is consistent with an approximately flat $\textnormal{SFR} - L_\textnormal{acc}$ relation, in all redshift bins. This is in agreement with some previous studies \citep[e.g][]{priddey03, shao10, dicken12, mullaney12, harrison12, rosario13, ma15, baner15, stanley15}, but not others \citep[e.g.][]{netzer09, hatzimi10, ima11, rafferty11, mull12a, chen13, Young2014, delv15, xu15, bian16}. 

Several explanations for an observed lack of a relation between SFR and $L_\textnormal{acc}$ have been proposed. They fall, broadly, into four categories. First, the strength of the relation depends on redshift, with an observable relation only emerging at $z\gtrsim2$ \citep{hatzimi10, rovilos12, delv15, harris16}. This could arise due to, for example, a higher free gas fraction at higher redshifts. Second, AGN luminosities can vary substantially on time-scales of days to months. If AGN luminosities are measured using methods that are particularly sensitive to such variations, such as X-ray observations \citep{barr86,ulr97,grupe01,chitnis09}, then underlying trends could be masked \citep{gabor13, volon15}. Third, it is possible that at very high SFRs and/or $L_\textnormal{acc}$ values, we are sampling an intrinsically different quasar population, one in which e.g. a different trigger for star formation leads to a weaker or absent correlation between SFR and $L_\textnormal{acc}$. Fourth, at high SFRs or high $L_\textnormal{acc}$, the properties of the starburst become decoupled from those of the AGN, plausibly because internal self-regulation processes become the dominant effect in regulating luminosities on time-scales comparable to or shorter than the AGN or starburst duty cycles.

The first possibility is, {\itshape on its own}, not plausible for our sample, as our sample spans a wide range in redshift and we do not see a significant correlation in any redshift bin. The second seems unlikely, since our sample is large and has AGN luminosities measured from optical, rather than X-ray data. We note however that there exists a conceptually similar possibility - that the duty cycles of starburst and AGN activity are sufficiently mismatched in duration that no correlation is observed between their luminosities. The available constraints on the duty cycle lengths of starburst and AGN phases are however not accurate enough for us to comment on this possibility. The third possibility also seems unlikely, since we found no differences between the {\it Herschel}/SDSS sample and the general SDSS quasar population in Section \ref{sec:sdss}, though we cannot formally exclude it.

The fourth scenario, however, is plausible, since our sample harbours the highest SFRs seen in quasars, where self-regulation may be expected to be seen, if it occurs. Moreover, it is consistent with observations of systems with comparable SFRs at $z>1$, i.e. submillimetre galaxies (SMGs, \citealt{barger14}), which exhibit a flattening in their SFRs at extremely high SFR values. An interesting corollary comes from the work of MY15, who find that star-forming regions in IR-luminous quasars do not increase in size at very high SFRs, but instead may increase in SFR {\itshape density}. An increase in SFR density could mean that starburst self-regulation becomes more effective, via e.g. approaching the Eddington limit for star formation, where winds from supernovae can expel free gas and thus stall star formation \citep[e.g.][]{thom05,murray05,dia12}. We thus propose that the flat relation between SFR and $L_\textnormal{acc}$ in our sample arises because self-regulation by the starburst becomes the dominant factor controlling the SFRs. We cannot, however, rule out a contribution from redshift-driven effects, or from mismatches in the lengths of duty cycles of starburst and AGN episodes.

The top panel of Fig. \ref{fig:SFR_trends} also shows no downturn in SFRs at the highest $L_\textnormal{acc}$ values. This is, in principle, inconsistent with the idea that luminous AGN can quench star formation in their host galaxies (see also e.g. \citealt{vil16}). An alternative interpretation is however possible; {\itshape assuming that AGN feedback occurs}, this instead suggests that $L_\textnormal{acc}$ (as derived from rest-frame UV observations) is not a good proxy for the strength of AGN feedback. This is consistent with AGN feedback being a brief phase in the AGN duty cycle, signposted by properties other than AGN luminosity \citep[e.g.][]{farrah12}.

\subsection{Star formation, black hole masses, and Eddington ratios}\label{sec:discsfagn2}
The middle panel of Fig. \ref{fig:SFR_trends} is consistent with there being no relation between SFR and $M_\textnormal{BH}$. This lack of a correlation is, in some senses, surprising. For example, it has been suggested that SFRs should scale with $M_\textnormal{BH}$ in quasars since SFRs should scale with the total available free gas mass or its large-scale distribution  \citep{cen15}, and both the free gas mass and $M_\textnormal{BH}$ should scale with the total baryon density. Thus, on $\gtrsim100$\,Myr time-scales, SFRs and $M_\textnormal{BH}$ in quasars may be expected to correlate since a larger free baryon reservoir would favor both higher current SFRs and higher $L_\textnormal{acc}$ {\itshape in the past}. We, however, do not see such a correlation.

A plausible explanation comes from considering our results in context with those of H16, who found that SFRs in quasars at $2<z<3$ do correlate with $M_\textnormal{BH}$ (and $L_\textnormal{acc}$), but only up to an SFR of $\sim600$\,M$_{\odot}$yr$^{-1}$. Above this SFR, they find no correlation. Our SFRs are mostly higher than those in H16 and are measured individually, rather than stacked as in H16, so it is interesting to compare their results to ours. In the top and middle panels of Fig. \ref{fig:SFR_trends}, we plot the relations that H16 derive between SFR and $L_\textnormal{acc}$ and $M_\textnormal{BH}$, respectively. For the SFR-$L_\textnormal{acc}$ plot our results are, given the sizes of the error bars on the average points in each redshift bin, consistent with the H16 relation. For the SFR-$M_\textnormal{BH}$ plot the consistency is weaker, but fig. 17 of H16 shows, plausibly, a flattening in the SFR-$M_\textnormal{BH}$ relation at about $600$\,M$_{\odot}$yr$^{-1}$ as well. Overall therefore, we argue that our results and those in H16 are both consistent with the idea that SFRs in quasars correlate with $L_\textnormal{acc}$ and $M_\textnormal{BH}$, but only up to an SFR when internal `self-regulation' by the starburst becomes important. 

The bottom panel of Fig. \ref{fig:SFR_trends} is consistent with there being no relation between SFR and $\lambda_\textnormal{Edd}$, at any redshift. This result is also in line with those in H16, who also find no evidence for an $\textnormal{SFR} - \lambda_\textnormal{Edd}$ relation, at any SFR (modulo possible enhanced SFRs at low $\lambda_\textnormal{Edd}$ values. Overall, this is consistent with there being no relation between SFRs in quasars and how efficiently the black hole is accreting, at {\itshape any} SFR. It is also consistent with the suggestion that $\lambda_\textnormal{Edd}$ values in quasars do not depend strongly on redshift, at least at $z\gtrsim0.5$ (e.g. \citealt{shan13,capl15}, see also \citealt{fine06,bluck11,luss12}).

\subsection{Star formation in broad absorption-line quasars }\label{sec:dischibal}
We find that there is no difference in the SFR, $L_\textnormal{acc}$, $M_\textnormal{BH}$, and $\lambda_\textnormal{Edd}$ distributions between the HiBAL and non-BAL quasars in our sample; all are consistent with being drawn from the same parent population. A minor caveat to this result is that our sample may be incomplete for HiBALs at very high redshifts (Sec. \ref{sec:sdss}), an effect that we cannot account for using available data; we think it unlikely that this incompleteness could be a significant factor leading to the lack of differences between BAL and non-BAL quasars that we observe.

This result aligns with previous studies, which find no differences between HiBAL and classical quasars, at any redshift \citep{priddey03,priddey07,gallagher07,pu15,harris16}. In particular, it is consistent with the study of \citealt{cao12}, who also use {\it Herschel}-SPIRE data to compare the properties of BAL versus non-BAL quasars, from the H-ATLAS survey, and find no differences between the two populations. The two studies, however, sample different regimes in SFR; our study examines quasars with typical SFRs of $\sim1000$\,M$_{\odot}$yr$^{-1}$, whereas the typical SFRs of the quasars in \citealt{cao12} are $\sim240$\,M$_{\odot}$yr$^{-1}$. The combined results suggest that BALs are not seen preferentially in certain $L_\textnormal{acc}$ or $M_\textnormal{BH}$ regimes, and not over specific ranges in SFR, even for those quasars harbouring the most luminous star formation events seen in any quasar at any redshift.

It has been argued that BAL quasars may be a promising quasar population within which to look for evidence of AGN feedback, since the BAL winds provide a natural mechanism to couple momentum from accretion-disk winds to ISM gas. The lack of any differences between the HiBALs and the non-BALs in our sample however argues that HiBAL quasars are, as a population, not sites for AGN feedback, unless the AGN feedback phase is much shorter than the lifetime of the BAL winds. Instead, our results are consistent with HiBAL quasars being normal quasars observed along a particular line of sight, with the outflows in HiBAL quasars not having any measurable effect on the star formation in their hosts (see also e.g. \citealt{vio16}).

\section*{ACKNOWLEDGEMENTS}
Funding for the creation and distribution of the SDSS Archive has been provided by the Alfred P. Sloan Foundation, the Participating Institutions, the National Aeronautics and Space Administration, the National Science Foundation, the U.S. Department of Energy, the Japanese Monbukagakusho and the Max Planck Society. The SDSS website is http://www.sdss.org/. The results described in this paper are based on observations obtained with {\it Herschel}, an ESA space observatory with science instruments provided by European-led Principal Investigator consortia and with important participation
from NASA. SPIRE has been developed by a consortium of institutes led by Cardiff Univ. (UK) and including: Univ. Lethbridge (Canada); NAOC (China); CEA, LAM (France); IFSI, Univ. Padua (Italy); IAC (Spain); Stockholm Observatory (Sweden); Imperial College London, RAL, UCL-MSSL, UKATC, Univ. Sussex (UK); and Caltech, JPL, NHSC, Univ. Colorado (USA). This development has been supported by national funding agencies: CSA (Canada); NAOC (China); CEA, CNES, CNRS (France); ASI (Italy); MCINN (Spain); SNSB (Sweden);
STFC, UKSA (UK); and NASA (USA). This research has made use of data from the HerMES project (http://hermes.sussex.ac.uk/). HerMES is a {\it Herschel} Key Programme utilizing Guaranteed Time from the SPIRE instrument team, ESAC scientists, and a mission scientist. HerMES is described in \cite{oliver12}. The HerMES data presented in this paper are available through the {\it Herschel} Database in Marseille (http://hedam.lam.fr/HerMES/).  AF acknowledges support from the ERC via an Advanced Grant under grant agreement no. 321323 – NEOGAL. This work makes extensive use of TOPCAT \citep{taylor05}.

\label{lastpage}

\end{document}